\documentclass[twocolumn,prl]{revtex4}
\usepackage{float}
\usepackage{amsfonts}
\usepackage{makeidx}
\usepackage{graphicx}
\usepackage{amsmath,amssymb,graphics,epsfig,color,times,bbm}

\begin{document}

\title{Graphical description of local Gaussian operations for continuous-variable weighted graph states}
\author{Jing Zhang$^{\dagger }$}
\affiliation{State Key Laboratory of Quantum Optics and Quantum
Optics Devices, Institute of Opto-Electronics, Shanxi University,
Taiyuan 030006, P.R.China}

\begin{abstract}
The form of a local Clifford (LC, also called local Gaussian (LG))
operation for the continuous-variable (CV) weighted graph states is
presented in this paper, which is the counterpart of the LC
operation of local complementation for qubit graph states. The novel
property of the CV weighted graph states is shown, which can be
expressed by the stabilizer formalism. It is distinctively different
from the qubit weighted graph states, which can not be expressed by
the stabilizer formalism. The corresponding graph rule, stated in
purely graph theoretical terms, is described, which completely
characterizes the evolution of CV weighted graph states under this
LC operation. This LC operation may be applied repeatedly on a CV
weighted graph state, which can generate the infinite LC equivalent
graph states of this graph state. This work is an important step to
characterize the LC equivalence class of CV weighted graph states.
\end{abstract}

\maketitle

Graph states \cite{one,two} - or equivalently called stabilizer
states, are special instances of multiparty quantum sates that are
of interest in a number of domains in quantum information theory and
quantum computation. Graph states can be defined in terms of the
stabilizer formalism, which is a group-theoretic framework
originally designed to construct broad classes of quantum
error-correcting codes - the stabilizer codes \cite{three}. In
addition to their role in quantum error-correction, graph states
have been used in a number of interesting applications, where the
measure-based model of quantum computation known as the one-way
quantum computer is certainly among the most prominent \cite{four}.

Most of the concepts of quantum information and computation have
been initially developed for discrete quantum variables, in
particular two-level or spin-$\frac{1}{2}$ quantum variables
(qubits). In parallel, quantum variables with a continuous spectrum,
have attracted a lot of interest and appear to yield very promising
perspectives concerning both experimental realizations and general
theoretical insights \cite{five,six}, due to relative simplicity and
high efficiency in the generation, manipulation, and detection of
continuous variable (CV) state. CV cluster and graph states have
been proposed \cite{seven}, which can be generated by squeezed state
and linear optics \cite{seven,eight,nine}, and demonstrated
experimentally for the four-mode cluster state \cite{ten,eleven}.
The one-way CV quantum computation was also proposed with the CV
cluster state \cite{twelve}. Moreover, the protocol of CV anyonic
statistics implemented with CV graph states is proposed
\cite{thirteen}.

It is well known that many graph states exhibit a high degree of
genuine multi-party entanglement \cite{forteen}, and that this
entanglement is a key ingredient responsible for the successful use
of these states in various applications. Therefore, a detailed study
of the entanglement properties of graph states is of natural
interest. The study of the nonlocal properties of graph states
naturally leads to an investigation of the action of local unitary
(LU) operations on graph states, and a classification of graph sates
under LU equivalence. Especially, a subclass of LU operations known
as local Clifford (LC) plays an important role. Due to the close
connection between the Pauli group, the stabilizer formalism and the
local Clifford group, the action of LC operation on graph states can
be described efficiently. Recently, the action of LC operations on
qubit graph states can entirely be understood in terms of a single
elementary graph transformation rule, called the local complement
rule \cite{fifteen,one}. A systematic classification of LC
equivalence of graph states has been executed \cite{one}. An
efficient algorithm (i.e., with polynomial time complexity in the
number of qubits) to decide whether two given stabilizer states are
LC equivalent, is known \cite{sixteen}. LU-LC equivalence problem
still was a long-standing open problem in quantum information
theory, which achieved the progress recently \cite{sixteen1}.

In the regime of continuous variable, LC equivalence of CV graph
states just began to be studied very recently. The local complement
rule was extended to the associated graphs of CV unweighted graph
states \cite{seventeen}. The simplest phenomenon was discussed
\cite{seventeen}, in which the corresponding LC operation was
presented for the local complementation on four-mode unweighted
graphs. It was shown that the corresponding LC operation for the
local complementation can not exactly mirror that for qubit, which
is not a single form compared with qubit. This result shows the
complexity of CV quantum systems. Whether the local complementation
for CV unweighted graphs can be implemented completely by the LC
transformations and the general form of the corresponding LC
operation can be found are still open question. In this paper, we
consider another way to investigate LC operation of CV graph states
as shown in Fig.1. First, the corresponding LC operation of local
complementation for qubit graph states is generalized to CV graph
states. Second, the CV weighted graph states is defined, which can
be expressed by the stabilizer formalism in terms of generators
within the Pauli group. It is distinctively different from the qubit
weighted graph states, which can not be expressed by the stabilizer
formalism \cite{forteen}. The action of this LC operation on the CV
weighted graph states is described by the graph rule. Thus, the
successive application of this LC operation can generate the LC
equivalence class of a CV weighted graph state with the infinite
elements. It is worth remarking that, whether the whole LC
equivalence class of a CV weighted graph state can be obtained by
repeatedly applying this LC operation, still need be further
investigated. In other words, what is the whole LC equivalence class
of a CV weighted graph state and how achieve it by LC operations?

%
\begin{figure}
\centerline{
\includegraphics[width=3in]{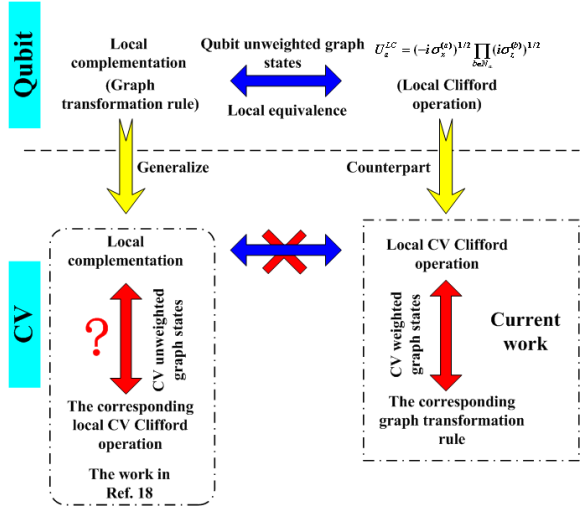}
} \vspace{0.1in}
\caption{(Color online). The diagram describing the LC equivalence
problem for qubit and CV graph states. \label{Fig1} }
\end{figure}

First, the CV operations \cite{eighteen} are presented briefly in
the follow. For CV, the Weyl-Heisenberg group, which is the group of
phase-space displacements, is a Lie group with generators
$\hat{x}=(\hat{a}+\hat{a}^\dagger)/\sqrt{2}$ (quadrature-amplitude
or position) and $\hat{p}=-i(\hat{a}-\hat{a}^\dagger)/\sqrt{2}$
(quadrature-phase or momentum) of the electromagnetic field. These
operators satisfy the canonical commutation relation
$[\hat{x},\hat{p}]=i$ (with $\hbar=1$). The single mode Pauli
operators are defined as $X(s)=exp[-is\hat{p}]$ and
$Z(t)=exp[it\hat{x}]$ with $s,t\in \mathbb{R}$. These operators are
non-commutative and obey the identity $ X(s)Z(t)=e^{-ist}Z(t)X(s)$.
In the Heisenberg picture, applying a Hamitonian $H$ gives a time
evolution for operators $\dot{A}=i[H,A]$, so that
$A(t)=exp[iHt]A(0)exp[-iHt]$. Accordingly, applying the Hamitonian
$H=\hat{x}$ for time $t$ takes $\hat{x}\rightarrow\hat{x},
\hat{p}\rightarrow\hat{p}-t$, and applying $H=-\hat{p}$ for time $s$
takes $\hat{x}\rightarrow\hat{x}-s, \hat{p}\rightarrow\hat{p}$. The
Pauli operator $X(s)$ is a position-translation operator, which acts
on the computational basis of position eigenstates as
$X(s)|q\rangle=|q+s\rangle$, whereas $Z$ is a momentum-translation
operator, which acts on the momentum eigenstates as
$Z(t)|p\rangle=|p+t\rangle$. The transformation of the Pauli
operators on the basis of position (momentum) eigenstates may be
derived as follows. Let $\hat{x}'= X(s)\hat{x}X(-s)=\hat{x}-s$, and
consider $\hat{x}'|q\rangle$. On the one hand, it must be
$\hat{x}'|q\rangle= (\hat{x}-s)|q\rangle= (q-s)|q\rangle$. On the
other hand, it also is $\hat{x}'|q\rangle=X(s)\hat{x}X(-s)|q\rangle=
X(s)\hat{x}|q-s\rangle=(q-s)X(s)|q-s\rangle =(q-s)|q\rangle$. Thus
$X(s)|q\rangle=|q+s\rangle$ is the correct operation. Similarly, it
may be shown that $Z(t)|p\rangle=|p+t\rangle$ is also the correct
transformation. The Pauli operators for one mode can be used to
construct a set of Pauli operators $\{X_{i}(s_{i}),Z_{i}(t_{i});
i=1,...,n\}$ for n-mode systems. This set generates the Pauli group
$\mathcal{C}_{1} $. The clifford group $\mathcal{C}_{2} $ is the
normalizer of the Pauli group, whose transformations acting by
conjugating, preserve the Pauli group $\mathcal{C}_{1} $; i.e., a
gate $\emph{U}$ is in the Clifford group if
$\emph{UR}\emph{U}^{-1}\in\mathcal{C}_{1}$ for every
$\emph{R}\in\mathcal{C}_{1}$. The Clifford group $\mathcal{C}_{2} $
for CV is shown \cite{eighteen} to be the (semidirect) product of
the Pauli group and linear symplectic group of all one-mode and
two-mode squeezing transformations. Transformation between the
position and momentum basis is given by the Fourier transform
operator $F=exp[i(\pi/4)(\hat{x}^{2}+\hat{p}^{2})]$, with
$F|q\rangle_{x}=|q\rangle_{p}$. This is the generalization of the
Hadamard gate for qubits. The phase gate $
P(\eta)=exp[i(\eta/2)\hat{x}^{2}]$ with $\eta\in \mathbb{R}$ is a
squeezing operation for CV and the action $P(\eta)RP^{-1}(\eta)$ on
the Pauli operators is
\begin{eqnarray}
P(\eta):X(s)&\rightarrow& e^{-is^{2}\eta/2}Z(s\eta)X(s)
,\nonumber \\
Z(t)&\rightarrow& Z(t),  \label{phase}
\end{eqnarray}
in analogy to the phase gate of qubit. The controlled operation C-Z
is generalized to controlled-$ Z (C_{Z})$. This gate
$C_{Z}=exp[i\hat{x}_{1}\bigotimes\hat{x}_{2}]$ provides the basic
interaction for two mode 1 and 2, and describes the quantum
nondemolition (QND) interaction. This set $ \{X(s), F, P(\eta), C-Z;
s,\eta \in \mathbb{R}\}$ generates the Clifford group. Here the
controlled operation with any interaction strength
$C_{Z}(\Omega)=exp[i\Omega\hat{x}_{1}\bigotimes\hat{x}_{2}]$
($\Omega\in \mathbb{R}$) will be used in the following. Another type
of the phase gate will also be utilized $
P_{X}(\eta)=FP(\eta)F^{-1}=exp[i(\eta/2)\hat{p}^{2}]$ and the action
on the Pauli operators is
\begin{eqnarray}
P_{X}(\eta):X(s)&\rightarrow& X(s)
,\nonumber \\
Z(t)&\rightarrow& e^{-it^{2}\eta/2}X(-t\eta)Z(t),  \label{phase1}
\end{eqnarray}
where $P_{X}(\eta)^{\dagger}=P_{X}(\eta)^{-1}=P_{X}(-\eta)$.

%
\begin{figure}
\centerline{
\includegraphics[width=3in]{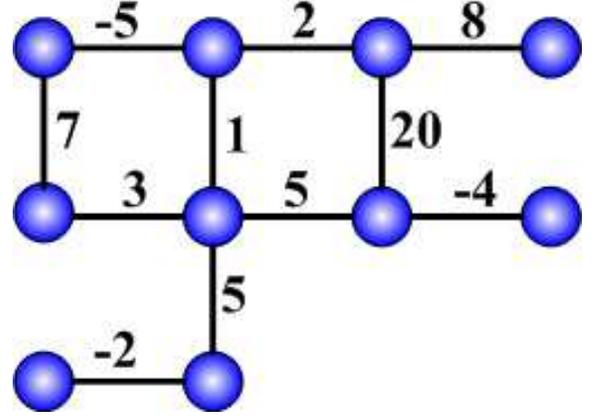}
} \vspace{0.1in}
\caption{(Color online). Example for depicting the CV weighted graph
state. \label{Fig2} }
\end{figure}

A weighted graph quantum state is described by a mathematical graph
$G=(V,E)$, i.e. a finite set of $n$ vertices $V$ connected by a set
of edges $E$ \cite{forteen}, in which every edge is specified by a
factor $\Omega_{ab}$ corresponding to the strength the modes a and b
have interacted as shown Fig.2. The preparation procedure of CV
weighted graph states is only to use the Clifford operations: first,
prepare each mode (or graph vertex) in a phase-squeezed state,
approximating a zero-phase eigenstate (analog of Pauli-X
eigenstates), then, apply the QND coupling ($C_{Z}(\Omega)$) with
the differen interaction strength $\Omega_{jk}$ to each pair of
modes $(j,k)$ linked by a weighted edge in the graph. Note that CV
unweighted graph states is to use the QND interaction all with the
same strength. Since all C-Z gates commute, the resulting CV graph
state becomes, in the limit of infinite squeezing,
$g_{a}=(\hat{p}_{a}-\sum_{b\in
N_{a}}\Omega_{ab}\hat{x}_{b})\rightarrow0$, where the modes $a\in V$
correspond to the vertices of the graph of $n$ modes, while the
modes $b\in N_{a}$ are the nearest neighbors of mode $a$. This
relation is as a simultaneous zero-eigenstate of the
position-momentum linear combination operators. The corresponding
$n$ independent stabilizers for CV weighted graph states are
expressed by $G_{a}(\xi)=exp[-i \xi g_{a}]=X_{a}(\xi)\prod_{b\in
N_{a}}Z_{b}(\Omega_{ab}\xi)$ with $\xi\in \mathbb{R}$. Note that it
is distinctively different from the qubit weighted graph states,
which can not be expressed by the stabilizer formalism
\cite{forteen}. The main reason induced this difference is that the
C-Z gate for qubit is periodic as a function of the interaction
strength, however, the CV C-Z gate is not.

%
\begin{figure}
\centerline{
\includegraphics[width=3in]{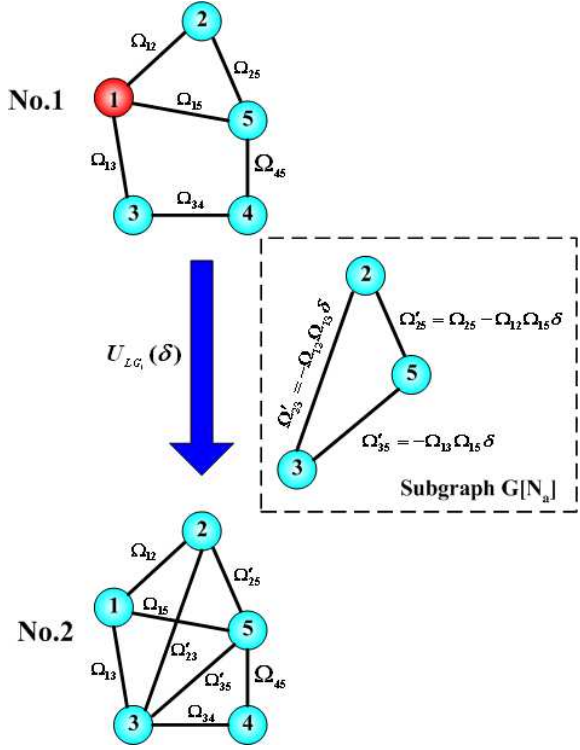}
} \vspace{0.1in}
\caption{(Color online). Example of the graph rule of the LG
operation $U_{LG_{1}}$ applied on a CV weighted graph state.
\label{Fig3} }
\end{figure}

The action of the local complement as the graph rule, can be
described as: letting $G=(V,E)$ be a graph and $a\in V$ be a vertex,
the local complement of $G$ for $a$, denoted by $\lambda_{a}(G)$, is
obtained by complementing the subgraph of $G$ generated by the
neighborhood $N_{a}$ of $a$ and leaving the rest of the graph
unchanged. The successive application of this rule suffices to
generate the complete orbit of any qubit graph states. The
corresponding LC operation of local complement for the qubit graph
states is a single and simple form, which is expressed by
$U_{a}^{LC}=(-i\sigma_{x}^{(a)})^{1/2}\prod_{b\in
N_{a}}(i\sigma_{z}^{(b)})^{1/2}$ \cite{fifteen,one}. This formalism
may be straightforward to generalize to CV weighted graph state,
which is expressed by
\begin{eqnarray}
U_{LG_{a}}(\delta)=P_{Xa}(-\delta)\prod_{b\in
N_{a}}P_{b}(\Omega_{ab}^{2}\delta). \label{LC}
\end{eqnarray}
Now the action of this LC operation on CV weighted graph states is
translated into transformations on their associated graphs, that is,
to derive transformations rules, stated in purely graph theoretical
terms, which completely characterize the evolution of CV weighted
graph states under this LC operation. The graph rule of applying
this LC operation is described as: first obtain the subgraph of $G$
generated by the neighborhood $N_{a}$ of $a$, then reset the weight
factor of all edges of this subgraph calculated with the equation
$\Omega_{b_{i}b_{j}}'=\Omega_{b_{i}b_{j}}-\Omega_{ab_{i}}\Omega_{ab_{j}}\delta$,
at last delete all the edges with the weight factor of zero, and
leave the rest of the graph unchanged. Here, a subgraph $G[C]$ of a
graph $G=(V,E)$, where $C\subset V$, is obtained by deleting all
vertices and the incident edges that are not contained in $C$.
Figure 3 presents an example of this graph rule applied on a CV
weighted graph state. The five independent stabilizers of the
weighted graph state No.1 $|\psi^{(1)}\rangle$ are given by
\begin{eqnarray}
G_{1}^{(1)}(\xi)&=&X_{1}(\xi)Z_{2}(\Omega_{12}\xi)Z_{3}(\Omega_{13}\xi)Z_{5}(\Omega_{15}\xi),\nonumber \\
G_{2}^{(1)}(\xi)&=&X_{2}(\xi)Z_{1}(\Omega_{12}\xi)Z_{5}(\Omega_{25}\xi),\nonumber
\\
G_{3}^{(1)}(\xi)&=&X_{3}(\xi)Z_{1}(\Omega_{13}\xi)Z_{4}(\Omega_{34}\xi),\nonumber
\\
G_{4}^{(1)}(\xi)&=&X_{4}(\xi)Z_{3}(\Omega_{34}\xi)Z_{5}(\Omega_{45}\xi),\nonumber
\\
G_{5}^{(1)}(\xi)&=&X_{5}(\xi)Z_{1}(\Omega_{15}\xi)Z_{2}(\Omega_{25}\xi)Z_{4}(\Omega_{45}\xi),\label{stab1}
\end{eqnarray}
with $G_{i}^{(1)}(\xi)|\psi^{(1)}\rangle=|\psi^{(1)}\rangle$ in the
limit of infinite squeezing, where $i=1,...,5$. Applying the LC
operation $U_{LG_{1}}(\delta)$ to the vertex 1, the five independent
stabilizers of the resulting graph state
$|\psi^{(2)}\rangle=U_{LG_{1}}(\delta)|\psi^{(1)}\rangle$ are
calculated by Eqs. \ref{phase},\ref{phase1},\ref{LC},\ref{stab1} and
with the relationship
$U_{LG_{1}}(\delta)G_{1}^{(1)}(\xi)=G_{1}^{(1)}(\xi)U_{LG_{1}}(\delta)$,
for example calculating $G_{2}^{(2)}(\xi)$,
\begin{eqnarray}
|\psi^{(2)}\rangle&=&U_{LG_{1}}(\delta)G_{2}^{(1)}(\xi)U_{LG_{1}}^{-1}(\delta)U_{LG_{1}}(\delta)|\psi^{(1)}\rangle\nonumber \\
&=&[e^{-i\xi^{2}\delta\Omega_{12}^{2}/2}Z_{2}(\delta\Omega_{12}^{2}\xi)X_{2}(\xi)]\times\nonumber\\&&
[e^{i(\Omega_{12}\xi)^{2}\delta/2}X_{1}(\delta\Omega_{12}\xi)Z_{1}(\Omega_{12}\xi)]\times\nonumber\\&&
Z_{5}(\Omega_{25}\xi)U_{LG_{1}}(\delta)|\psi^{(1)}\rangle\nonumber \\
&=&X_{2}(\xi)Z_{1}(\Omega_{12}\xi)Z_{5}(\Omega_{25}\xi)[X_{1}(\delta\Omega_{12}\xi)Z_{2}(\delta\Omega_{12}^{2}\xi)]\times\nonumber\\&&
U_{LG_{1}}(\delta)G_{1}^{(1)}(-\delta\Omega_{12}\xi)|\psi^{(1)}\rangle\nonumber \\
&=&X_{2}(\xi)Z_{1}(\Omega_{12}\xi)Z_{3}(-\Omega_{12}\Omega_{13}\delta\xi)\times\nonumber\\&&
Z_{5}((\Omega_{25}-\Omega_{12}\Omega_{15}\delta)\xi)|\psi^{(2)}\rangle\nonumber \\
&=&G_{2}^{(2)}(\xi)|\psi^{(2)}\rangle
\end{eqnarray}
to obtain
\begin{eqnarray}
G_{1}^{(2)}(\xi)&=&X_{1}(\xi)Z_{2}(\Omega_{12}\xi)Z_{3}(\Omega_{13}\xi)Z_{5}(\Omega_{15}\xi),\nonumber \\
G_{2}^{(2)}(\xi)&=&X_{2}(\xi)Z_{1}(\Omega_{12}\xi)Z_{3}(-\Omega_{12}\Omega_{13}\delta\xi)\nonumber\\&&
Z_{5}((\Omega_{25}-\Omega_{12}\Omega_{15}\delta)\xi)\nonumber \\
&=&X_{2}(\xi)Z_{1}(\Omega_{12}\xi)Z_{3}(\Omega_{23}'\xi)Z_{5}(\Omega_{25}'\xi),\nonumber
\\
G_{3}^{(2)}(\xi)&=&X_{3}(\xi)Z_{1}(\Omega_{13}\xi)Z_{2}(\Omega_{23}'\xi)Z_{4}(\Omega_{34}\xi)Z_{5}(\Omega_{35}'\xi),\nonumber
\\
G_{4}^{(2)}(\xi)&=&X_{4}(\xi)Z_{3}(\Omega_{34}\xi)Z_{5}(\Omega_{45}\xi),\nonumber
\\
G_{5}^{(2)}(\xi)&=&X_{5}(\xi)Z_{1}(\Omega_{15}\xi)Z_{2}(\Omega_{25}'\xi)Z_{3}(\Omega_{35}'\xi)Z_{4}(\Omega_{45}\xi).\label{stab2}
\end{eqnarray}
which exactly correspond to the stabilizers of No.2 weighed graph
state in Fig.3.

%
\begin{figure}
\centerline{
\includegraphics[width=3in]{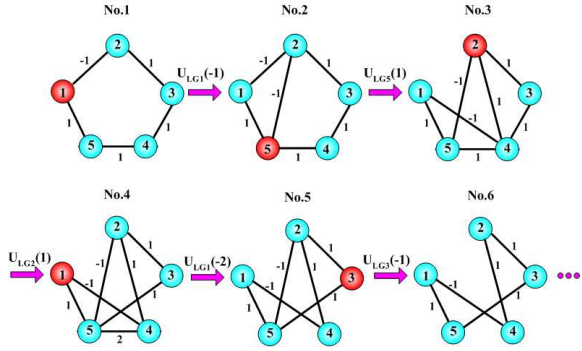}
} \vspace{0.1in}
\caption{(Color online). Example of the graph rule of the LG
operation $U_{LG_{a}}$ repeatedly applied on a CV weighted graph
state. The rule is successively applied to the vertex, which is
colored with red in the figure. \label{Fig4} }
\end{figure}

This LC operation may be applied repeatedly on a CV weighted graph
state, which can generate the LC equivalence class of this graph
state. Figure 4 shows an example of how to repeatedly apply this
rule to obtain the LC equivalence class of a CV weighted graph
state. Note that the elements in the LC equivalence class of a CV
weighted graph state, generated by the LC operation $U_{LG_{a}}$,
are infinite, and whether the whole LC equivalence class of a CV
weighted graph state can be obtained by repeatedly applying this LC
operation, still need be further investigated.

%
\begin{figure}
\centerline{
\includegraphics[width=3in]{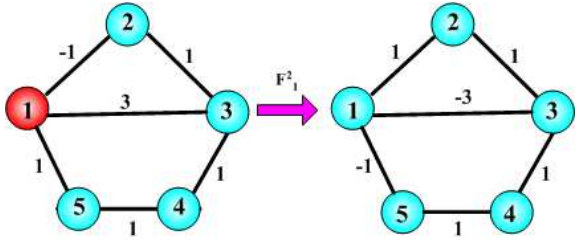}
} \vspace{0.1in}
\caption{(Color online). Example of the graph rule of the LG
operation $F^{2}$ applied on a CV weighted graph state.
\label{Fig4} }
\end{figure}

At last, the graph rules of two extra and very useful LC operations
are presented. One of the LC operations is $F^{2}$, corresponding to
the square of the Fourier transform operator, which is used in
Ref.\cite{seventeen}. This operation has the effect of taking
$F^{2}\hat{x}(F^{2})^{-1}=-\hat{x}$ and
$F^{2}\hat{p}(F^{2})^{-1}=-\hat{p}$. The graph rule of applying this
LC operation $F^{2}$ on a vertex a is described as: add the negative
sign on the weight factor of all edges connecting the vertex a. An
example for the LC operation $F^{2}$ is shown in Fig.5. The other LC
operation is $S(r)=exp[ir(\hat{x}\hat{p}+\hat{p}\hat{x})/2]$ with
$r\in \mathbb{R}$, which is a quadrature squeezing operation for CV
corresponding to the the phase-sensitive optical parametric
amplifier. The action of $S(r)$ on the position and momentum
operators is $S(r)\hat{x}S(r)^{-1}=\hat{x}e^{r}$ and
$S(r)\hat{p}S(r))^{-1}=\hat{p}e^{-r}$, which means to stretch the
position component and squeeze the momentum component of an optical
field. The graph rule of applying this LC operation $S(r)$ on a
vertex a is described as: multiply $e^{-r}$ on the weight factor of
all edges connecting the vertex a. An example for the LC operation
$S(r)$ is shown in Fig.6. Note that whether these two LC operations
are the necessary transformations for the LC equivalence of CV
weighted graph states, still need be further studied.

%
\begin{figure}
\centerline{
\includegraphics[width=3in]{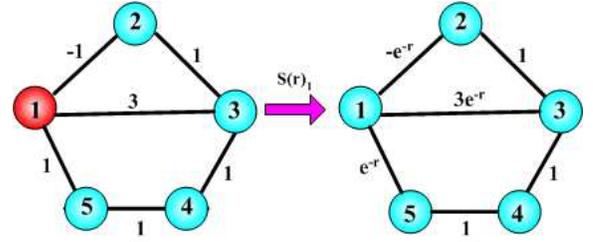}
} \vspace{0.1in}
\caption{(Color online). Example of the graph rule of the LG
operation $S(r)$ applied on a CV weighted graph state. \label{Fig4}
}
\end{figure}

In summary, the corresponding LC operation of local complementation
for qubit graph states is extended to CV weighted graph states. This
LC operation may be applied repeatedly on a CV weighted graph state,
which can generate the local Clifford equivalence class of this
graph state with the infinite elements. This work is an important
step to characterize the LC equivalence class of CV weighted graph
states. It is natural to raise the question with this work whether a
polynomial time algorithm can be found to decide whether two CV
graph states are LG equivalent and the action of local Gaussian
group on CV graph states can be translated into elementary graph
transformations characterized by several simple rules just like
qubit graph states. Furthermore, LU equivalence for CV graph states,
which is same as that for qubit graph states, also is an open
problem.

$^{\dagger} $Corresponding author's email address:
jzhang74@sxu.edu.cn, jzhang74@yahoo.com

\section{\textbf{ACKNOWLEDGMENTS}}

J. Zhang thanks K. Peng and C. Xie for the helpful discussions. This
research was supported in part by NSFC for Distinguished Young
Scholars (Grant No. 10725416), National Basic Research Program of
China (Grant No. 2006CB921101), NSFC Project for Excellent Research
Team (Grant No. 60821004), NSFC (Grant No. 60678029), Program for
the Top Young and Middle-aged Innovative Talents of Higher Learning
Institutions of Shanxi and NSF of Shanxi Province (Grant No.
2006011003).

\end{document}